\documentclass{aastex}
\usepackage{emulateapj5,apjfonts,psfig}

\def\asca{{\sl ASCA }}

\def\ros{{\sl ROSAT }}

\def\ein{{\sl Einstein }}

\def\chandra{{\sl Chandra }}

\def\it{\sl}

\slugcomment{accepted for publication in The Astrophysical Journal Letters}

\shorttitle{X-RAY EMISSION LINES FROM $\theta^1$ Ori C}

\shortauthors{SCHULZ et al.}

\begin{document}

\title{X-ray Line Emission from the Hot Stellar Wind of $\theta^1$ Ori C}
\author{
N. S. Schulz,
C. R. Canizares,
D. Huenemoerder,
and
J.C. Lee
 }
\affil{Center for Space Research, Massachusetts Institute of Technology,
Cambridge, MA 02139.}

\begin{abstract}
We present a first emission line analysis of a high resolution X-ray spectrum of the stellar wind 
of $\theta^1$ Ori C
obtained with the High Energy Transmission grating Spectrometer onboard the Chandra X-ray Observatory.
The spectra are resolved into a large number of emission lines from H- and He-like O, Ne, Mg, Si, S, Ar and
Fe ions. The He-like Fe XXV and 
Li-like Fe XXIV appear quite strong indicating very hot emitting regions. From H/He flux ratios, as well as from
Fe He/Li emission measure ratios we deduce  temperatures ranging from 0.5 to 6.1$\times 10^7$ K.
The He-triplets are very sensitive to density as well. At these temperatures the relative strengths of
the intercombination and forbidden lines indicate electron densities well above 10$^{12}$ cm$^{-3}$.
The lines appear significantly broadened from which we deduce a mean velocity of 770 km s$^{-1}$
with a spread between 400 and 2000 km s$^{-1}$. Along with results of the deduced emission measure we conclude
that the X-ray emission could originate in dense and hot regions with a characteristic size of less
then 4$\times 10^{10}$ cm.
\end{abstract}

\keywords{
stars: early type ---
X-rays: stars ---
binaries: stellar winds ---
techniques: spectroscopic}

\section{Introduction}

X-ray spectra from stellar winds
have been quite difficult to interpret, since the first observations with \ein (Seward et al. 1979). 
The spectral resolution from previous X-ray missions was insufficient for resolving
any emission and absorption lines. 
Models that describe
the mass loss from a hot luminous star successfully in the UV (Pauldrach et al. 1994, Lamers et al.
1999) could never correctly predict observed X-ray fluxes.
A model involving a possible hot corona seems unlikely due to the lack evidence
for absorption at the 20.6 \AA~ K-shell
ionization edge from oxygen (Cassinelli $\&$ Swank 1983); additionally no coronal line emission
(i.e. Fe XIV at 530 nm) was found in optical spectra of $\eta$ Ori and $\kappa$ Ori (Nordsieck et al. 1981)
nor in the protoptype star for stellar winds $\zeta$ Pup (Baade $\&$ Lucy (1987).   
X-ray emission from shocks emerging in instabilities within a 
radiatively driven wind forming a forward shock was proposed by Lucy $\&$ White (1980) and  Lucy (1982). However,
reverse shocks that decelerate material as it rams into dense shells ahead
seem more likely (Owocki et al. 1988). These models predict X-rays up to a temperature of about 
kT$\sim$0.5 keV. Feldmeier et al. (1997) introduced turbulent perturbations into their calculations
which ultimatively allowed for higher temperatures. 

Several recent observations with the  0.1 to 10.0 keV \asca bandpass, 
i.e. of $\tau$ Sco (Cohen et al. 1997) and $\eta$ Car (Tsuboi et al. 1997), 
indicated that there is a hard 
component to these spectra at $>$ 2 keV. 
Although these two examples tend to resemble quite extreme cases
of massive stars, \asca observations of other, less massive O-stars (Corcoran et al. 1994),
and the Orion Trapezium (Yamauchi et al. 1996) indicated the existence of a similar hard component in
their spectra as well. 
In the most recent analysis of \chandra
CCD spectra of the Orion Trapezium, Schulz et al. (2000) were able to resolve the entire Trapezium into
individual sources. Identified O- and B-stars showed a soft component at $\sim$ 0.8 keV as observed with 
\ros; the very early spectral types required an additional  hard component of 
temperatures above 2 keV as well. 

In this letter we present first results from an observation of $\theta^1$ Ori C with the High Energy Transmission
Grating Spectrometer (HETGS, Canizares et al. 2000, in preparation) onboard \chandra (Weisskopf et al. 1996). 
We present an X-ray line list of the brightest emission lines detected and
investigate emission line properties to determine the range of temperatures and densities of the line emitting regions.
Some of the results will have immediate impact constraining models of stellar winds. 

\centerline{\epsfxsize=7.5cm\epsfbox{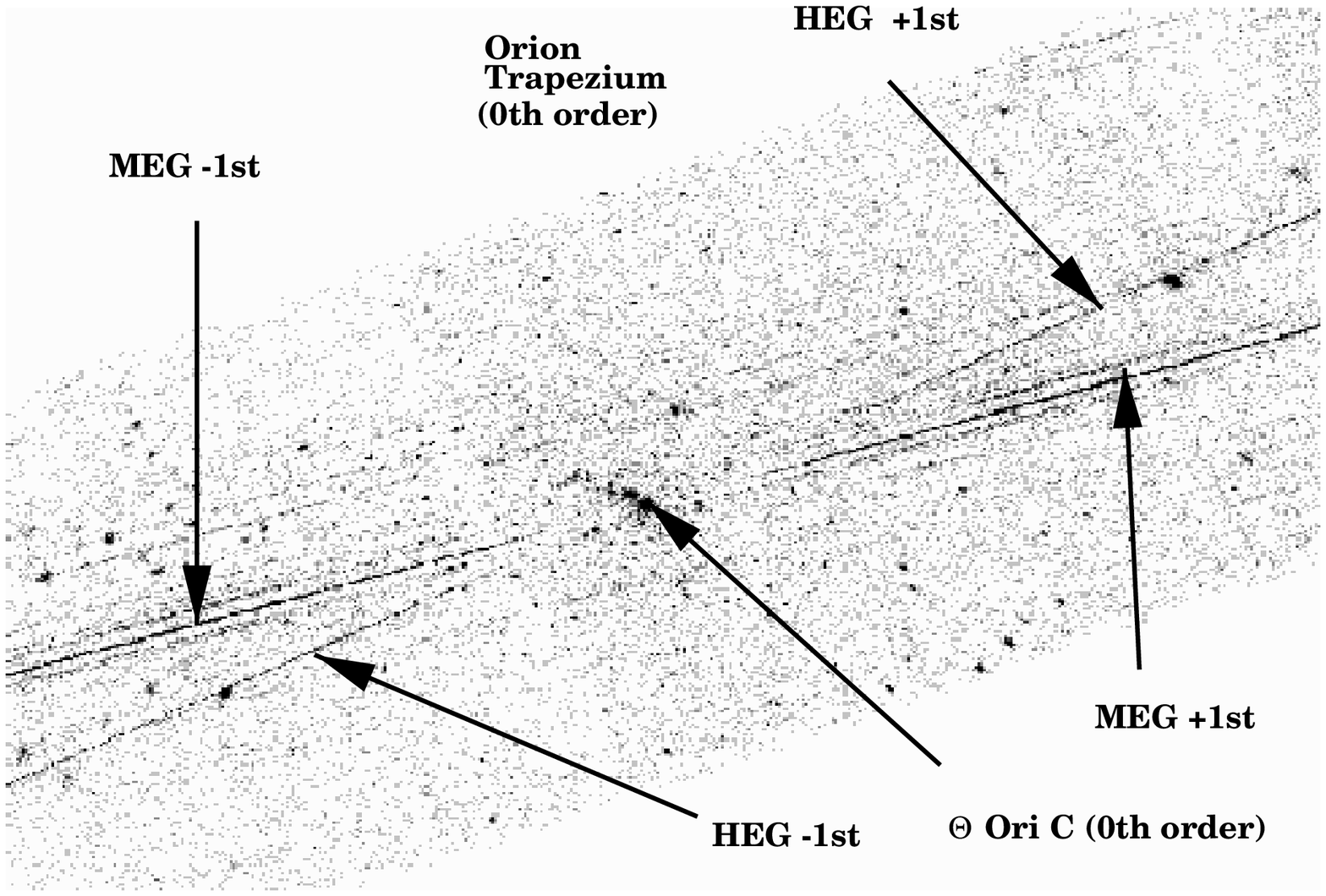}}
\figcaption{HETGS focal plane image of the Orion Trapezium Cluster
\label{focal}}
   
\section{Chandra Observations and Data Reduction}

$\theta^1$ Ori C was observed with the HETGS
on 1999 October 31 (05:47:21 UT) continuously for 53 ks.
The HETGS carries two different types of transmission
gratings, the Medium Energy Gratings (MEG) and the High Energy Gratings (HEG) with
grating constants of 4000 and 2000 \AA, respectively. It allows for high resolution spectroscopy between
1 and 35 \AA ~with a peak spectral resolution at 12 \AA ~of $\lambda/\Delta\lambda \sim 1400$
and at 1.8 \AA ~of $\lambda/\Delta\lambda \sim 180$ in 1st order HEG. 
The dispersed spectra were recorded with 
the Advanced CCD Imaging Spectrometer (Garmire et al. 2000, in preparation). 
We also refer to the available Chandra X-ray Center (CXC) docments for more detailed descriptions.

We recorded
a total of 4.5$\times10^4$ events in the co-added 1st order MEG and 1.7$\times10^4$ events in the HEG
after standard grade selection.
The level 1 event lists provided by the CXC were reprocessed 
using the latest available data processing input products.
One of the major difficulties in the spectra extraction was that $\theta^1$ Ori C is located
in the core of the Orion Trapezium Cluster, which has several sources that are bright enough to produce
dispersed spectra as well as many X-ray sources in the vicinity
which then would imitate emission lines by coincidence. Most of these contributions can be
removed by checking for dislocations of the source point spread functions from the center of the
cross-dispersion profile, by cross-correlating with each dispersion arm,
and finally by the energy discrimination of the focal plane CCDs.  

In order to determine the zero point of the wavelength scale we  
fitted the dispersed images of the MEG and HEG and determined the intersections of these
fits with the zero order read out trace of the CCD. The fits of the MEG and HEG were consistent within
0.3 detector pixels ($\sim$ 0.003 \AA ~in MEG and 0.002 \AA ~in HEG
1st order). The current status of the overall wavelength calibration is of the order of 0.1$\%$ and mostly
depends on uncertainties in the position of CCD gaps ($\pm$ 0.5 pixels). 
This leads to a worst case uncertainty in the scale of 0.01 \AA ~in 1st order MEG, 0.008 \AA ~in 1st order HEG,
0.003 \AA ~in 3rd order MEG. Level 1 to 1.5 event list processing as well as aspect corrected exposure map
computations were done  
using available CXC software, the final grating spectra were extracted using custom software and FTOOLs,
s pectral fits were performed with ISIS. 

\section{HETGS Spectra}

\begin{figure*}
\centerline{\epsfxsize=18.5cm\epsfysize=5.8cm\epsfbox{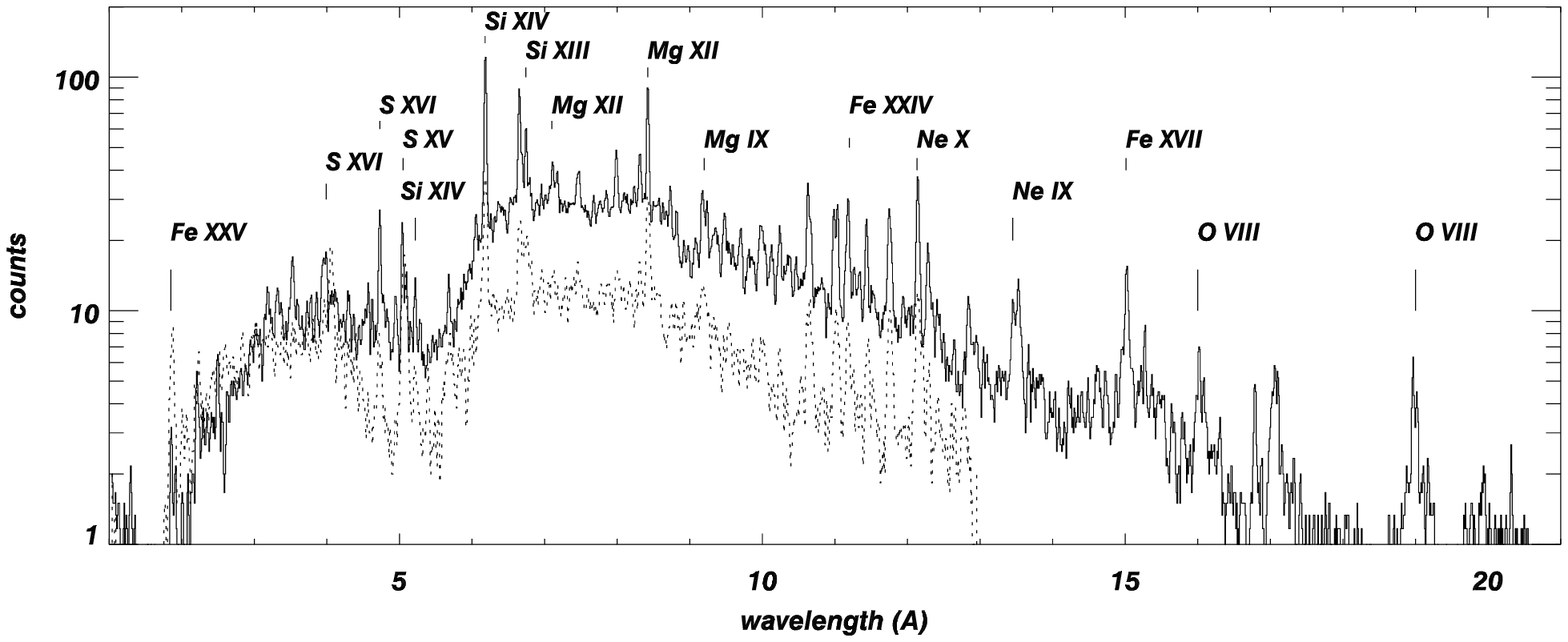}}
\figcaption{The 1st order MEG (solid line) and HEG (dotted line) count spectra
\label{counts}}
\end{figure*}

Figure 1 shows the focal plane image of the Orion Trapezium Cluster and its dispersed spectra. Several sources
in the cluster are bright enough to produce highly resolved grating spectra. The brightest trace in the middle
corresponds to $\theta^1$ Ori C; arrows point to its 0th order image as well as the -1st and +1st MEG and HEG
spectral trace.  
The extracted and cleaned spectra are shown in Figure 2. The solid line shows the MEG co-added 1st order spectrum,
similarly the dotted line for the HEG. The spectral binning is 0.01 \AA~.
The spectra show a substantial continuum and numerous emission lines.
At the current stage of analysis we concentrate on relevant properties of the emission lines;
global model fits will be left for subsequent papers.

\subsection{Emission Lines}

Table 1 displays prominent and for this analysis most interesting set of lines we have identified so far.
We observe lines in a band
from 1.8 to 19 \AA~. There is
not much flux above 22 \AA~, since $\theta^1$ Ori C is moderately absorbed with a
column density of 1.93$\times 10^{21}$ cm$^{-2}$ (Schulz et al. 2000).
The detected lines are from all major abundant elements, i.e from H-like and He-like ions of O, Ne, Mg,
Si, S, and Fe. Calcium and Argon lines are not apparent - they are either missing or below
our statistical detection limits. The Lyman-$\alpha$
lines are strongly detected in all
other species but Fe. For some ions we also
observe the Lyman-$\beta$ and higher transitions. In 5 ions, S, Si, Mg, Ne, and O,
we resolve the He-like triplets into resonance (r), intercombination (i), and forbidden (f) lines.
The Ne triplet is contaminated by Fe transitions.
The triplet in Fe XXV is not resolved and the line appears considerably broadened.
Iron is quite abundant and in addition to many
lower transitions, we observe strong lines from He-, Li-, and Be-like Fe.
Here we only include Fe XXV and some lines from Fe XXIV and Fe XVII, which we use for plasma diagnostics,
a full investigation of the Fe lines is under way.

\vbox{
\footnotesize
\begin{center}
{\sc TABLE 1\\
EMISSION LINES FROM H- AND HE-LIKE IONS AND FROM SOME FE XVII AND FE XXIV IONS.}
\vskip 4pt
\begin{tabular}{lcccccc}
\tableline
\tableline
     Ion & type & $\lambda (exp.)$ & $\lambda (meas.)^1$ & Flux$^2$ & Fwhm\\
         &  &\AA & \AA & & 10$^{-2}$\AA \\
\tableline
     & & & & \\
     Fe XXV  & He4 & 1.851 & 1.856 & 7.533 & 3.07$^{~4}$\\
     Fe XXIV & Li2 & 7.987 & 7.984 & 2.596 & 3.17\\
     S XVI   & H2  & 3.990 & 3.997 & 1.458 & 2.31 $^{~4}$\\
     S XVI   & H1  & 4.730 & 4.729 & 6.141 & 1.46 \\
     S XV    & He4 & 5.040 & 5.040 & 9.457 & 3.54 \\
     S XV    & He5 & 5.060 & 5.060 & 2.734 & 0.13 \\
     S XV    & He6 & 5.100 & 5.105 & 4.334 & 2.17 \\
     Si XIV  & H2  & 5.220 & 5.218 & 3.913 & 3.34 \\
     Si XIV  & H1  & 6.180 & 6.182 & 11.742& 0.81 \\
     Si XIII & He4 & 6.650 & 6.650 & 7.313 & 1.11 \\
     Si XIII & He5 & 6.690 & 6.690 & 4.068 & 1.64 \\
     Si XIII & He6 & 6.740 & 6.740 & 5.579 & 1.63$^{~4}$ \\
     Mg XII  & H2  & 7.100 & 7.105 & 3.218 & 2.34 \\
     Mg XI   & He2 & 7.470 & 7.466 & 2.100 & 2.51\\
     Mg XII  & H1  & 8.420 & 8.418 & 8.755 & 1.38 \\
     Mg XI   & He4 & 9.170 & 9.169 & 3.000 & 1.65 \\
     Mg XI   & He5 & 9.230 & 9.237 & 1.000 & 0.24 \\
     Mg XI   & He6 & 9.310 & 9.316 & 0.600 &  --  \\
     Ne X    & H4  & 9.480 & 9.474 & 5.834 &  --  \\
     Fe XXIV & Li6A&11.031 &11.036 & 6.255 & 1.36\\
     Fe XXIV & Li6B&11.184 &11.178 & 5.879 & 0.35$^{~4}$\\
     Fe XXIV & Li7B&11.269 &11.267 & 1.539 & 1.65\\
     Fe XXIV & Li7A&11.445 &11.430 & 2.724 & 3.16\\
     Ne X    & H1  &12.130 &12.145 & 12.950& 3.47$^{~4}$ \\
     Ne IX   & He4 &13.440 &13.445 &  2.768& 0.71\\
     Fe XVII & Ne4 &15.013 &15.020 & 12.424& 4.59\\
     O VIII  & H2  &16.010 &16.019 &  2.441& 4.63$^{~4}$\\
     O VIII  & H1  &18.970 &18.991 &  9.547& 1.39\\
     O VII   & He4 &21.600 &21.600 &  1.428& 0.24\\
     O VII   & He5 &21.800 &21.790 &  2.452& 0.19\\
     & & & & \\
\tableline
\end{tabular}
\parbox{2.7in}{
\small\baselineskip 3pt
\footnotesize
\indent
$\rm ^1$ in units of $10^{-5}$ photons cm$^{-2}s^{-1}$ \\
$\rm ^2$ appear blended\\
}
\end{center}
\setcounter{table}{1}
\normalsize
\centerline{}
}

In order to determine line positions, widths and fluxes we treated the underlying continuum
as background and preformed local fits using a generic flat model for the background and
Gaussian functions for the lines. 
For the line identifications and nomenclature we use the line list 
presented by Mewe et al. (1985). 
The wavelength identification is based on the closest known transition within the instrument resolution.
Because of its superior resolution line centroids below 13 \AA~ are determined using the HEG. 
The difference between the measured and expected
line positions are well within the current uncertainties of the Chandra wavelength calibration;
no obvious systematic shifts other than a trend for larger deviations at higher wavelength exists,
which can be explained by the increasing uncertainty due to more CCD gaps in the dispersion direction.

\subsection{Velocities}

We mostly rely on the a factor 2 superior resolution of the HEG for line width measurements $<$ 13 \AA~.
These widths are shown in column 6 of Table 1. 
Instrumental corrections have been applied using the latest HETGS response files.
The most reliable widths come from the bright and unblended Lyman $\alpha$ lines, where we consider the fine structure
splits negligibly small. 

The FWHM of the lines range between 0.3 and 4.1$\times10^{-2}$ \AA~, 
an order of magnitude too large to be interpreted as thermal broadening.
If we deduce Doppler velocities from $\Delta \lambda / \lambda = v_D / c$, we
obtain a distribution of velocities with a mean value of 771 km s$^{-1}$ (Figure 3).
The Fe XXV width at 1.85 \AA~ appears exceptionally wide, but we do not have the sensitivity to  
resolve its r, i, and f components. However if we force the fit using the triplet model,
we obtain a mean width corresponding to  with a velocity of 1650 km s$^{-1}$.
Near 11 \AA~ we also introduce systematic
uncertainties by fitting weak as well as more blended lines adding to an additional scatter in the 
observed line widths. The drop at 11 \AA~, in that respect, may not be real. When we for the time being
ignore these weaker data points, the velocity scatter is $\sim$ 400 to 2000 km s$^{-1}$.  
We do not observe any  measureable asymmetries in the line shapes, which would indicate deviations
from an isotropic wind emission geometry.

\subsection{Temperatures $\&$ Electron Densities}

Flux ratios of highly ionized ions are sensitive to the density and temperature of a collisionally ionized
plasma (see Mewe 1999 for a review). We have several ways to estimate the local temperature from the
line. We can use the (f+i)/r flux ratio of the He-like triplet, flux ratio of
the H-like Lyman $\alpha$  to the He-like resonance lines, as well as He- to Li-like and
He- to Ne-like ion flux ratios.
Comparing the line ratios with theory is difficult and depends on a variety of assumptions.
Here we use the Astrophysical Plasma Emission Code (APEC) and the corresponding database
(APED \footnote{http://hea-www.harvard.edu/APEC/}) described by Smith et al. (1998).
In general we make the
assumption that for a specific transition the emitting plasma is isothermal or at least has
a symmetric distibution with that peak temperature. We also assume that the emitting plasma
is in ionization equillibrium.
The ratios within the He-like triplets are good indicators for the local plasma temperature.
A comparison of the flux ratios to calculated ratios from the database yields temperatures
of 4.2$\pm3.1\times 10^5$ K for O, 9.3$\pm3.8\times 10^6$ K for Mg, 2.0$\pm1.4\times 10^6$ K for Si,
and 1.5$\pm1.1\times 10^7$ K for S.     
The line flux ratios of the H- to He-like ions as well as the He- to Li-like Fe ions are good
indicators of the temperature largely indepedent of density. 
Here the ratios for the Lyman $\alpha$ transitions of O, Ne, Mg, Si, and S 
yields temperatures of 0.6$\pm$0.2, 0.7$\pm$0.2, 1.5$\pm$0.2, 2.0$\pm$0.3, and 2.3$\pm 0.3 \times 10^7$ K, 
respectively. The O, Ne, and Mg ratios have also been corrected for the effect of interstellar column density.
Since we do not detect the Lyman $\alpha$ line from Fe, we can only set an upper limit
to the temperature of 6.3$\times 10^7$ K by computing the ratio of its 3$\sigma$ detection limit to the 
whole detected He-like Fe triplet. This limit is consistent with the results from the
Li-like Fe-lines which predict temperatures between 3.5 and 5.1$\times 10^7$ K. 
From the the ratio with the Ne-like Fe XVII line we deduce a lower temperature
of 5.0$\times 10^6$ K.

\centerline{\epsfxsize=7.5cm\epsfbox{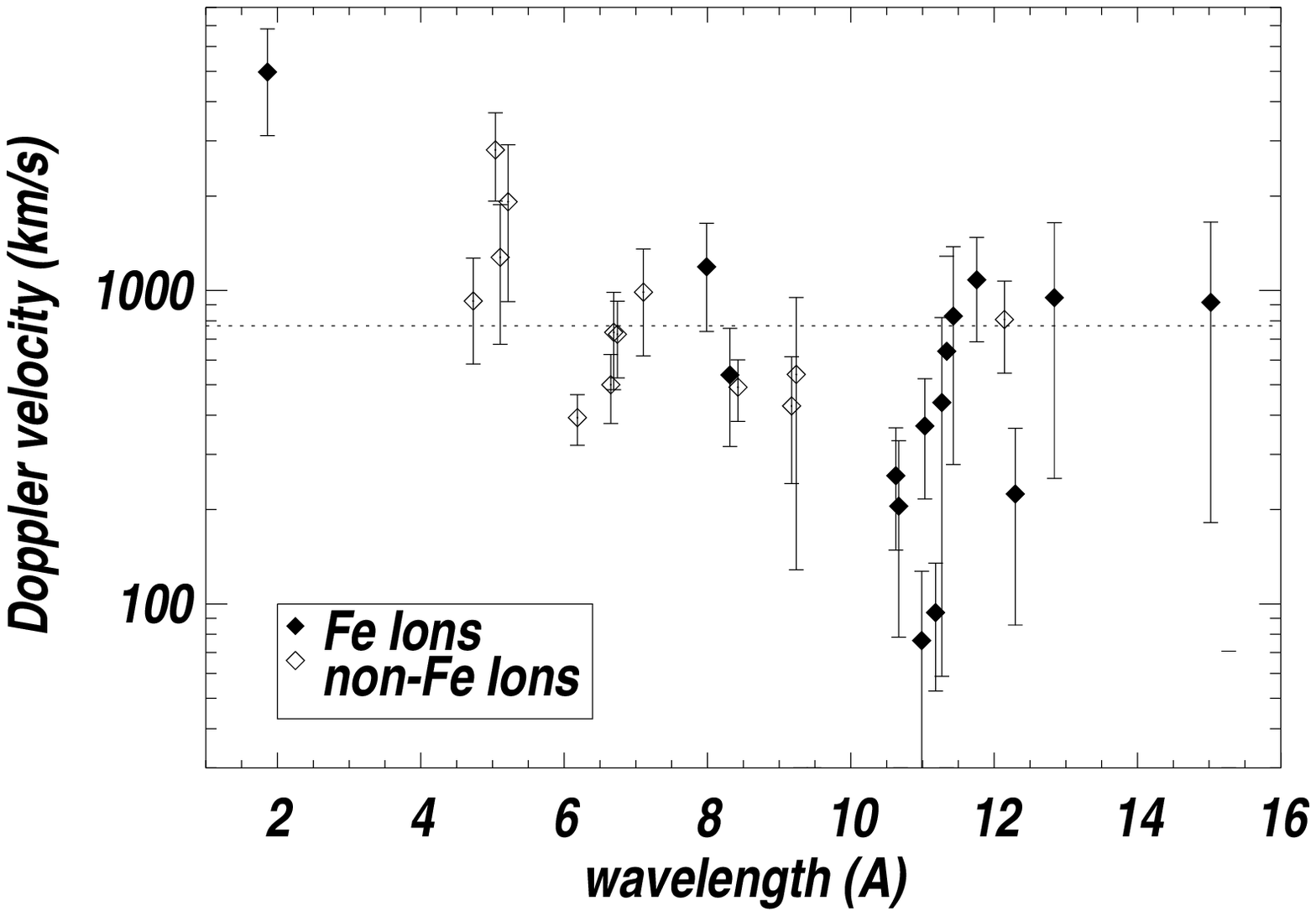}}
\figcaption{Instrument corrected Doppler velocities deduced from fitted line widths.
The dotted line shows the mean velocity, the dashed line the instrument width of the
HEG.
\label{velocities}}

Although the uncertainties in the (f+i)/r flux ratios are higher, since the ratio
uses 3 relatively weak lines, the temperatures for O and Si are significantly lower than
the ones deduced from the hydrogenic lines. In the case of Si this in part could be due contamination
of the forbidden line with the Mg XII Lyman $\gamma$ line. However, if we assume that we 
do not observe the Si XIII forbidden line at all, we deduce a temperature, which is still a factor 2 lower 
than the one from the hydrogenic lines. The fact that these temperatures are systematically
lower could either mean that we are not in ionization equillibrium, or, which we consider more likely, that part
of the He-like and lower transitions originate from different, somewhat cooler zones.  

The forbidden line intensity in the 
He-like triplets is metastable and thus density sensitive. However, it should also be noted
that this ratio may also be affected by external radiation fields (Kahn et al. 2000). Here we 
simply compare the
f/i ratio to the expectation from emissivity calculations from the APED database. 
In general, the ratios from elements of higher Z than Si are not suited for this analysis,
because here the ratio is not much sensitive to densities below $10^{15}$ cm$^{-3}$, which we
do not expect here.
From the database we deduce a lower density sensitivity limit of the f/i ratio to roughly
$10^{10}$ cm$^{-3}$ for O, $10^{12}$ cm$^{-3}$ for Mg, and $10^{13}$ cm$^{-3}$ for Si.    
For these ions we measure f/i ratios of 0.04$\pm$0.02, 0.60$\pm$0.18, and 1.37$\pm$0.51 corresponding
to electron densities to 3.5$\pm2.7\times 10^{12}$ cm$^{-3}$ for
oxygen, 4.3$\pm2.3\times 10^{13}$ cm$^{-3}$ for Mg, and 9.0$\pm10.6\times 10^{13}$ cm$^{-3}$ for Si.
We fixed the temperature to the ones obtained from the hydrogenic lines, because the latter
lines are less density sensitive and we also assume that they are not affected by opacity effects. 
The uncertainties reflect the average upper and lower density limits according to
the errors in the ratios.
The O-ratio is a lower limit, because as a result of not observing the forbidden
line, we estimated the flux based on a 3$\sigma$ detection limit.   
We observe a significant flux in the Mg XII Lyman $\beta$ line, which makes it very likely
that the forbidden line in Si at 6.74 \AA~ is contaminated with a Mg XII Lyman $\gamma$ line. From
the APED database we calculated a emissivity ratio of the Mg XII Lyman $\beta$ to Lyman $\gamma$
of 3.05. Based on the observed flux for the Lyman $\beta$ line we can then estimate the 
flux contribution of the Mg XII Lyman $\gamma$ line to the Si XIII forbidden line. 
The f/i ratio in Si reduces then from 1.37 to 1.11. 

Using the lines in Table 1 we also deduced volume emission measures (VEM) assuming 
isothermal and isotropic conditions. We obtain the lowest VEM from Fe XVII with 1.52$\times 10^{54}$ cm$^{-3}$
and the highest from S XVI with 2.5$\times 10^{56}$ cm$^{-3}$. If we apply the deduced
range of densities above and spherical 
geometry we find characteristic emission sizes between 4.0$\times10^{8}$ cm and 3.9$\times10^{10}$ cm. 

\section{Discussion}

A large number of emission lines were detected with a signal to noise ratio
of greater than 5$\sigma$. These lines are from H- and He-like ion species
of Fe, S, Si, Mg, Ne, and O; Ca and Ar are not detected.
We do not observe any residual line shifts we could attribute to the source.

The X-ray lines appear
significantly broadened implying a mean shock velocity of 770 km s$^{-1}$. 
A similar shock velocity of 500 km s$^{-1}$ has been determined from \ros fits of $\zeta$ Pup
spectra (Hillier et al. 1993). 
Within the statistical uncertainities and the current state of the analysis, 
the line profiles appear symmetric showing blue and red-shifts of equal proportions.
We therefore conclude that the wind emission is quite isotropic.
Our estimates from the VEM indicates that the emission volume is equivalent to
a volume of well within one stellar radius. However, Simulations by Feldmeier et al. (1997)
indicated that the X-ray emission must extend far out into the wind. Estimating
the distance of the X-ray producing shell collision we use the $\beta$ velocity law
(with $\beta$= 0.88) and a terminal velocity of $\sim$ 1000 km s$^{-1}$ (Prinja et al. 1990)
to determine the distance where the wind reaches 770 km s$^{-1}$,
which gives about 4 stellar radii. In this respect it is plausible that the X-ray emitting regions appear
in very thin dense shells as described by Feldmeier et al. (1997). Their model, however,
predicts strong variability of the X-ray emission, which, at least in terms of total
X-ray flux, is not observed during our observations. 

Whatever the actual emission
geometry turns out to be, the emitting regions of $\theta^1$ Ori C appear well confined, hot, and 
dense with a temperatures ranging between 0.05 and 6 $\times 10^7$ K (0.06 to 4.5 keV)
and densities above $10^{12}$ cm$^{-3}$.        
CCD spectra of $\theta^1$ Ori C already indicated that the X-ray emission must have
a hot component with a temperature of kT$\sim$3 keV (Yamauchi et al. 1997, Schulz et al. 2000)
or equivalent to $\sim$ 4$\times 10^7$ K. A broadband study ranging from UV to \asca spectra of the
B0 V star $\tau$ Sco Cohen (1996) also suggested that wide range of temperatures up to $10^7$ K is necessary to 
describe X-ray and UV emission. 
It is quite hard to explain such high temperature with the line driven shock instability
models (Lucy $\&$ White 1980, Lucy 1982), which predict temperatures more on the order of
5$\times 10^6$ K. The spectrum in Figure 2 also shows that the hard part of the 
spectrum cannot be explained by inverse Compton scattering of UV photons (Chen $\&$ White 1991),
since it shows emission lines with similar characteristics than the low energy lines.
The approach by Feldmeier et al. (1997) seems to provide promising ingredients to
describe the co-existence of low and high temperatures at relatively high densities.   
A most recent analysis 
by Kahn et al. (2000) of XMM RGS spectra of $\zeta$ Pup 
also resulted in low f/i line ratios and thus high density values, but it was suggested that
a high UV radiation field could destroy the He-like forbidden line leading to an overestimation
of the density.      

It still remains to be explained whether the high temperatures in $\theta^1$ Ori C
and other hot candidates are the exception or the rule. Clearly, we need more high resolution X-ray observations
of massive O-Stars in order classify the X-rays in terms of their line emission properties.
These properties include temperature ranges, densities, emission volumes and ionization
balance.    

\acknowledgments
The authors want to thank N. Brickhouse for providing additional atomic data tables
and the Chandra X-ray Center for its enormous support.
This research is funded by contracts SV-61010 and NAS8-39073.

\end{document}